\documentclass[11pt]{article}
\usepackage{menuproc}
\usepackage{psfig}
\begin{document}
\titlematter{$N^*$ Masses from QCD Sum Rules}
{Xinyu Liu$^a$ and Frank X. Lee$^{a,b}$}
{$^a$Center for Nuclear Studies, Department of Physics,\\ 
The George Washington University, Washington, DC 20052, U.S.A. \\
$^b$Jefferson Lab, 12000 Jefferson Avenue, Newport News, VA 23606, USA}
{We report $N^*$ masses in the spin-1/2 and spin-3/2 sectors
using the method of QCD Sum Rules.
They are based on three independent sets derived from generalized 
interpolating fields.
The predictive ability of each sum rule is examined by a 
Monte-Carlo based analysis procedure in which all three phenomenological 
parameters (mass, coupling, threshold) are extracted simultaneously.
A parity projection technique is also studied.}

%\section{Introduction}
The QCD Sum Rule method~\cite{SVZ79} is a time-honored method
that has proven useful in revealing a connection between 
hadron phenomenology and the non-perturbative nature of the QCD vacuum via 
only a few parameters (the vacuum condensates).
It has been successfully applied to a variety of observables in
hadron phenomenology, providing valuable insights from a
unique, QCD-based perspective, and continues an active field 
(try a keyword search with 'QCD Sum Rule').
The method is analytical (no path integrals!), is physically transparent (one can trace back term by term which operators are responsible for what),
and has minimal model dependence (Borel transform, and a continuum threshold).
The accuracy of the approach is limited due to limitations inherent 
in the operator-product-expansion (OPE), but well understood.

Our goal is to explore the possibility of using the method to understand 
the N* spectrum.  
The calculation of baryon masses in the approach is 
not new~\cite{Ioffe81,Bely82,Chung84,RRY85,Derek90}.
Here we focus on the excited states and emphasize the predictive ability of 
the method for N* properties based on careful analysis, using 
a rigorous Monte Carlo-based~\cite{Derek97} numerical analysis procedure 
that treats all three phenomenological parameters (mass, coupling, threshold) 
as free parameters and extracts them simultaneously with error bars.
In particular, we study the low-lying states in the spin-1/2 and 
spin-3/2 sectors with both positive and negative parity.
A similar analysis for the baryon decuplet has been done~\cite{Lee98}.

The starting point is the time-ordered, two-point correlation function in the 
QCD vacuum:
\begin{equation}
\Pi(p)=i\int d^4x\; e^{ip\cdot x}\langle 0\,|\,
T\{\;\eta(x)\, \bar{\eta}(0)\;\}\,|\,0\rangle,
\label{cf2pt}
\end{equation}
where $\eta$ is the interpolating field that has the 
quantum numbers of the baryon under consideration.
We consider the most general current for the nucleon with 
$I(J^P)={1\over 2}\left({1\over 2}^+\right)$,
\begin{equation}
\eta^{ N}_{1/2}(x)= -2\;\left[\,
\epsilon_{abc} \left(u^{aT}(x)C\gamma_5 d^b(x)\right)u^c(x)
+ \beta\; 
\epsilon_{abc} \left(u^{aT}(x)C d^b(x)\right)\gamma_5 u^c(x)
\,\right].
\end{equation}
Here $C$ is the charge conjugation operator, the superscript $T$ means transpose,
and $\epsilon_{abc}$ makes it color-singlet. 
The real parameter $\beta$ can be varied to achieve maximal overlap with  
the state in question.
The choice advocated by Ioffe~\cite{Ioffe81} and often used in QCD sum rules studies 
corresponds to $\beta=-1.0$. 
It is well-known that a baryon interpolating field couples to states of both parities, 
despite having an explicit parity by construction.
The results below will show that $\beta$ can be varied to saturate a sum rule 
with either positive or negative parity states.
For states with $I(J^P)={1\over 2}\left({3\over 2}^+\right)$,
we consider 
\begin{equation}
\eta^{ N}_{3/2,\mu}(x)=\epsilon_{abc}\left[
\left(u^{aT}(x)C\sigma_{\rho\lambda} d^b(x)\right)
\sigma^{\rho\lambda}\gamma_\mu u^c(x)
- \left(u^{aT}(x)C\sigma_{\rho\lambda} u^b(x)\right)
\sigma^{\rho\lambda}\gamma_\mu d^c(x) \right].
\end{equation}
The interpolating fields for $\Sigma$, $\Lambda$ and $\Xi$ can be obtained 
by appropriate substitutions of quark fields under SU(3) color symmetry or
flavor symmetry.

With two kinds of interpolating fields, 
three possible correlation functions can be constructed:
the correlator of generalized spin-1/2 currents 
$\eta_{1/2}$ and $\bar{\eta}_{1/2}$,
the mixed correlator of generalized spin-1/2 current 
$\eta_{1/2, \mu} =\gamma_\mu\gamma_5\, \eta_{1/2}$
and the spin-3/2 current $\bar{\eta}_{3/2, \nu}$,
and the correlator of spin-3/2 currents 
$\eta_{3/2, \mu}$ and $\bar{\eta}_{3/2, \nu}$.
From them, 11 independent sum rules emerge which can be used to study   
$1/2\pm$ and $3/2\pm$ states.

Table~\ref{tab33} shows the predictions for $1/2+$ states from 
the chiral-odd sum rules at the tenser structure $\gamma_\mu p_\nu \hat{p}$,
using the Monte-Carlo analysis. 
Sum rules fall into two categories: one with odd-dimension operators (chiral-odd)
and the other with even-dimension operators (chiral-even).
The predictions compare favorably with the observed values, with an accuracy 
of about 100 MeV. The couplings come as by-products which are useful in 
the calculation of matrix elements because they enter as normalization. 
Table~\ref{tab34} shows the predictions for $3/2-$ states.

\begin{figure}
\centerline{\psfig{file=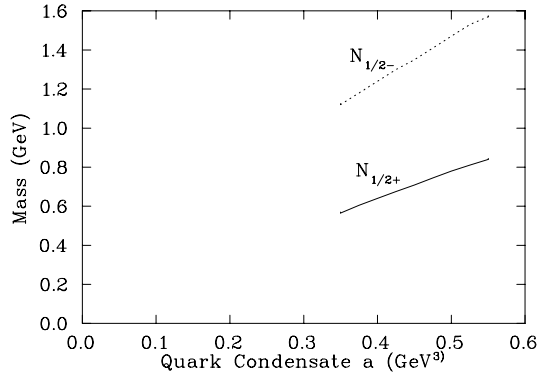,width=7.0cm,angle=90}}
\caption{Mass splitting between $N^*_{{1\over2}-}$ and
$N_{{1\over2}+}$ as a function of the quark condensate parameter 
 $a=-(2\pi)^2\,\langle\bar{q}q\rangle$. The physical point corresponds to 
$a=0.52 \mbox{ GeV}^3$.}
\label{split}
\end{figure}

One drawback in the conventional approach is that states with both parities 
contribute in the sum rules. Although sometimes one can saturate a sum rule 
with a certain parity by adjusting $\beta$, as done above, it is desirable to 
separate the two parities exactly.
This can be achieved by replacing the time-ordering operator $T$ 
in the correlation function in Eq.~(\ref{cf2pt}) with $x_0>0$, 
%
%\begin{equation}
%\Pi(p)=i\int d^4x\; e^{ip\cdot x}\,\theta(x_0)\,\langle 0\,|\,
%\eta(x)\, \bar{\eta}(0)\,|\,0\rangle,
%\end{equation}
%
and constructing sum rules in the complex $p_0$-space 
in the rest frame ($\vec{p}=0$)~\cite{Jido96}.
This is equivalent to a parity projection technique used in lattice QCD 
calculation of N* masses~\cite{zhou01}.
Table~\ref{tab35} shows the predictions for $1/2-$ states in this method.
The results are much improved, as indicated by the smaller error bars and 
very wide Borel regions. The agreement with experiment is excellent.
To further investigate the origin of splittings between parity partners, 
we show in Figure~\ref{split} the mass splittings 
between $N_{{1\over2}-}$ and $N_{{1\over2}+}$ as a function of 
the quark condensate (the order parameter of spontaneous chiral symmetry breaking).
One can see a clear decrease in the splitting with decreasing quark condensate,
in the range that the sum rule does not break down.

%\section{Conclusion}
In conclusion, we demonstrated the predictive power of QCD sum rules for 
N* masses in the low-lying $1/2\pm$ and $3/2-$ sectors, 
with an accuracy on the order of 5 to 10\%.
The parity separation method is promising. 
We are extending it to the spin-3/2 sector.
More analysis is under way to understand the details of the splitting patterns 
across particle channels and parities, 
%in terms of vacuum condensates and quark masses.
in terms of explicit and dynamical chiral symmetry breaking.
 
\acknowledgments{
This work is supported in part by U.S. Department of Energy under Grant DE-FG03-93DR-40774.}

\begin{table}
\begin{center}
\caption{Predictions for $1/2+$ states from the chiral-odd sum rules at the structure
$\gamma_\mu p_\nu \hat{p}$.}
\label{tab33}
\begin{tabular}{ccccccc}
\hline \\
Sum Rule & Region & $w$ & $\tilde{\lambda}_{1/2} \tilde{\alpha}_{1/2}$ & Mass &
Exp.\\
& (GeV) &  (GeV) & (GeV$^6$) & (GeV) & (GeV)\\ \hline
$N_{{1\over 2}+} \, (\beta=+1.0)$
& 1.06 to 1.46  & 1.31$\pm$ .22 & 1.13$\pm$ .53 & 1.06$\pm$ .11 &
0.938 \\
$\Sigma_{{1\over 2}+} \, (\beta=+1.0) $
& 1.12 to 1.53  & 1.48$\pm$ .23 & 1.62$\pm$ 0.68 & 1.16 $\pm$ .12 & 1.193
\\
$\Xi_{{1\over 2}+} \, (\beta=+1.0)$
& 1.35 to 1.80 & 1.69$\pm$ .27  & 2.61$\pm$ 1.20 & 1.32$\pm$ .14 & 1.318
\\
$\Lambda_{{1\over 2}+} \, (\beta=+1.0)$
& 1.28 to 1.72 &  1.53$\pm$ .24  & 0.66$\pm$ 0.28 & 1.23$\pm$ .12 & 1.116 \\
\hline
\end{tabular}
\end{center}
\end{table}
\begin{table}
\begin{center}
\caption{Predictions for $3/2-$ states from the chiral-odd sum rules 
at the structure $g_{\mu\nu}$.}
\label{tab34}
\begin{tabular}{ccccccc}
\hline \\
Sum Rule & Region & $w$ & $\tilde{\lambda}_{3/2}^2$ & Mass &
Exp.\\
& (GeV) &  (GeV) & (GeV$^6$) & (GeV) & (GeV)\\ \hline
$N_{{3\over 2}-}$
& 0.95 to 1.17  & 1.65$\pm$ .24 & 27.6$\pm$ 11.8 & 1.44$\pm$ .13 &
1.520 \\
$\Sigma_{{3\over 2}-}$
& 1.29 to 1.36  & 1.91$\pm$ .25 & 46.6$\pm$ 20.1 & 1.69 $\pm$ .14 & 1.580
\\
$\Xi_{{3\over 2}-}$
& 1.30 to 1.39 & 2.19$\pm$ .27  & 84.8$\pm$ 42.9 & 1.84$\pm$ .16 & 1.820
\\
$\Lambda_{{3\over 2}-}$
& 1.22 to 1.32 & 2.01$\pm$ .25  & 19.8$\pm$ 9.3 & 1.71$\pm$ .15 & 1.690 \\
\hline 
\end{tabular}
\end{center}
\end{table}
\begin{table}
\begin{center}
\caption{Predictions for $1/2-$ states from the new method where 
parity is exactly separated.}
\label{tab35}
\begin{tabular}{ccccccc}
\hline \\
Sum Rule & Region & $w$ & $\tilde{\lambda}_{3/2}^2$ & Mass &
Exp.\\
& (GeV) &  (GeV) & (GeV$^6$) & (GeV) & (GeV)\\ \hline
$N_{{1\over 2}-} \, (\beta=+1.1) $
& 0.80 to 1.80  & 2.26$\pm$.08 & 4.58$\pm$.38 & 1.53$\pm$.05 &
1.535 \\
$\Sigma_{{1\over 2}-}  \, (\beta=+1.1)$
& 0.80 to 1.90  & 2.35$\pm$.06 & 5.74$\pm$.69 & 1.63$\pm$.07 & 1.620 \\
$\Xi_{{1\over 2}-}  \, (\beta=+1.1)$
& 0.80 to 1.80 & 2.38$\pm$.06  & 5.42$\pm$.68 & 1.61$\pm$.08 & 1.620 \\
$\Lambda_{{1\over 2}-}  \, (\beta=+1.1)$
& 0.90 to 1.80 & 2.49$\pm$.04  & 7.06$\pm$.55 & 1.67$\pm$.05 & 1.670 \\
\hline 
\end{tabular}
\end{center}
\end{table}

\end{document}